# Interaction of Epithelial Cells with Surfaces and Surfaces Decorated by Molecules

D. Martini[1], O. Marti[1], M. Beil[2], T. Paust[1], C. Huang[3], M. Moosmann[3], J. Jin[3], T. Heiler[3], R. Gröger[3], T. Schimmel[3], S. Walheim[3]

[1] Institute for Experimental Physics, Ulm University, 89069 Ulm
[2] Clinic for Internal Medicine, Ulm University, 89069 Ulm
[3] Institute for Applied Physics and Institute of Nanotechnology,
  Karlsruhe Institute of Technology, 76128 Karlsruhe

## Summary

*A detailed understanding of the interface between living cells and substrate materials is of rising importance in many fields of medicine, biology and biotechnology.* Cells at interfaces often form epithelia. The *physical barrier* that they form is one of their main functions. It is governed by the properties of the networks forming the cytoskeleton systems and by cell-to-cell contacts. Different substrates with varying surface properties modify the *migration velocity of the cells*. On the one hand one can change the materials composition. Organic and inorganic materials induce differing migration velocities in the same cell system. Within the same class of materials, a change of the surface stiffness or of the surface energy modifies the migration velocity, too.

For our cell adhesion studies a variety of different, homogeneous substrates were used (polymers, bio-polymers, metals, oxides).

In addition, an effective lithographic method, *Polymer Blend Lithography (PBL)*, is reported, to produce patterned Self-Assembled Monolayers (SAM) on solid substrates featuring two or three different chemical functionalities. This we achieve without the use of conventional lithography like e-beam- or UV lithography, only by using self-organization. These surfaces are decorated with a Teflon-like and with an amino-functionalized molecular layer. The resulting pattern is a copy of a previously created self-organized polymer pattern, featuring a scalable lateral domain size in the sub-micron range down below 100 nanometers. The resulting monolayer pattern features a high chemical and biofunctional contrast with feature sizes in the range of cell adhesion complexes like e.g. focal adhesion points.

*The interaction of the cells with the substrate is mediated by cell-boundary crossing matrices of macromolecular processes*. The cells optimize their complexes such as to adapt their mechanical properties to the substrate. These *mechanical properties are probed by force-distance curves measured on living cells* and by scanning force microscopy of them. Microrheology of extracted cytoskeletons complemented the data on the mechanics of cytoskeletons. We concentrated our investigations on cells from the pancreas carcinoma and on some cell lines for comparison. *Videomicroscopy was used to measure the substrate dependent motility*. We found characteristic differences in the migration velocity as well as hints to correlations to the cell mechanics. *These quantities can be used to characterize cells.*

## 1  Introduction

To form a barrier is one oft the principal functions of epithelial tissues. On a cellular level this function depends among other factors on the properties of the networks forming the cytoskeleton and the cell-to-cell contacts. A modulation of the tissue function can be achieved by structuring the substrate [1]. A detailed understanding of the interface between cells and substrates is of importance in medicine, in biology and in biotechnology. The morphological and functional integrity of epithelia is regulated by receptor-mediated interactions with the environment, e.g. extracellular matrix proteins. The in-vitro engineering of epithelial tissues is based on providing guidance for individual cells to migrate to and adhere at pre-defined locations. This can be achieved by modulating the molecular structure of artificial surfaces [1]. Both organic as well as inorganic molecules have been used for this purpose [2]. This study investigated the impact of various materials on the motile properties of epithelial cells. By using different materials, cells of different origins can be provided with environmental clues to migrate into their designated positions during tissue engineering.

In this work we investigate the cell motion on various substrates by video microscopy. The elastic properties of living cells were measured by extracting Force-Distance-Curves. The values, dependent on the substrate, were correlated with the motion of these cells on the substrates. Microrheology allows a measurement of the cytoskeleton mechanical properties alone. Here we concentrated on the keratin cytoskeleton. Furthermore patterned substrates were produced by Lithography with the aim to have custom tailored substrates for cell adhesion studies. It is proposed that the substrate-dependent motilities could be useful for directing cells to a human designed arrangement. Especially the use of patterned substrates including micro- and nanostructured substrates and patterns presenting gradients in size and concentration of bio-functional areas could lead to a pattern-guided locomotion of cells, which opens perspectives for future tissue engineering with a tailored architecture consisting of different cell types.

## 2  Results and Discussion

### 2.1  Patterning Substrates with Polymer Blend Lithography

A rapid and cost-effective lithographic method, *Polymer Blend Lithography* (PBL), is reported to produce patterned Self-Assembled Monolayers (SAM) on solid substrates featuring two or three different chemical functionalities [3]. For the pattern generation we use the phase separation of two immiscible polymers in a blend solution during a spin coating process. By controlling the spin coating parameters and conditions, including the ambient atmosphere (humidity), the molar mass of the polystyrene (PS) and poly-methyl-methacrylate (PMMA) and the mass ratio between the two polymers in the blend solution, the formation of a purely lateral morphology (PS islands standing on the substrate while isolated in the PMMA matrix) can be reproducibly induced. Either of the formed phases (PS or PMMA) can be selectively dissolved afterwards and the remaining phase can be used as a lift-off mask for the formation of a nanopatterned functional silane monolayer. This "monolayer copy" of the polymer phase morphology has a topography of about 1.3 nm. A demonstration of PS islands diameter tuning is given by changing the molar mass of PS. Moreover, polymer blend lithography can provide the possibility of fabricating a surface with three different chemical components: This is demonstrated by inducing breath figures (evaporated condensed entity) at higher humidity during the spin-coating process. Here we demonstrate the formation of a lateral pattern consisting of regions covered with perfluordecyltrichlorsilan (FDTS), amino-propyltriethoxy-silane (APTES) and at the same time featuring regions of bare $SiO_x$. The patterning process could be applied even on meter-sized substrates with various functional SAM molecules, making this process suitable for the rapid preparation of quasi two-dimensional nanopatterned functional substrates e.g. for the template-controlled growth of ZnO nanostructures [4].

**Introduction:** Self-assembled monolayers (SAMs) are well-known and have been intensively studied for many years – partly because of their interesting properties and partly because of interesting perspectives for potential applications as functional, ultrathin coatings [5–8]. Due to their functionality SAMs play an important role for the construction of sensors [9,10] or e.g. the controlling of cell adhesion [11]. Patterning of self-assembled monolayers on the nanometer scale is easily performed by sequential lithographic techniques well established in literature. Electron beam lithography allows to desorb or to destroy molecules of a SAM layer line by line [12,13]. Advanced Scanning Force Microscopy (SFM) techniques allow not only to image the topography of surfaces but also the spatially resolved study of surface properties such as electrical, elastic, tribological and wear properties [14–26]. At the same time, scanning force microscopy-based lithographic techniques allow the structuring and patterning of surfaces with a lateral resolution down to the nanometer scale [27–33]. The advantage of techniques such as electron beam lithography or SFM-based lithography is their high lateral resolution and their reproducibility – their major disadvantage is the fact that they rely on sequential writing processes which are very time consuming and need expensive equipment. For patterning larger areas on the nanometer scale – e.g. for the fabrication of nanopatterned, biofunctional templates – easy-to-use, cheap and fast techniques allowing the parallel fabrication of billions of nanostructures are required.

Phase separation of binary polymer blend solutions during a spin-coating process produces nano- and micropatterns on large areas in a fast and upscalable fashion. This phase separation has been intensively studied over the past two decades and allows the formation of complex layered or lateral micro- or nanoscale structures [34–41]. These structures can be used for many applications such as anti-reflection coatings [42], photovoltaic devices [43,44], organic light emitting diodes (OLED) [45–47] and more. Polymer phase separation in thin films can be obtained by methods such as spin-coating [34] and Langmuir-Schaefer deposition [48]. In the case of the spin-coating technique it is possible to guide the morphogenesis by a pre-patterned solid template in order to form layout-defined structures [49–51]. However, so far there is no direct way to use the resulting polymer blend film as a lithographic mask because the formed structure contains both lateral and layered phase separations [52–54]. Special techniques, such as UV curing have to be combined to get the film ready for lithographic applications [55,56]. Zemla et al. [55] describe a technique where after cross-linking one polymer, the other one is removed and a protein is adsorbed at the free surface areas. The second polymer, however, cannot be dissolved due to the cross-linking and remains on the substrate. In Ref. [56] Kawamura et al. use the difference of resistance against photo etching between the two polymers in the blend to remove the component with less stability under photo radiation. The remaining micro-pattered polymer layer has a thickness of about 3 nm however, without well-defined surface chemistry.

Here, we are aiming for a lateral polymer phase morphology which can be completely removed by a selective solvent to make the substrate available for well-defined chemical surface modification. This can be achieved by inserting a silane SAM which then exposes a functional group. The preparation process of the SAM should not affect the remaining polymer mask so it can protect the substrate during the procedure and can be removed afterwards. For spin coating of polymer blend films, there are many parameters and conditions, such as the concentration of the polymer solution, the spin rate, the surface property of the substrate and more that affect the final morphology of the polymer blend film. Some examples of both the influence of the substrate [57–59] and the solution parameters [60–63] can be found in the recent literature. We found that the formed polymer blend structures in our case are also strongly dependent upon the relative humidity during the demixing. The relative humidity influences the interaction of the two polymer phases and the affinity of the polymers to the substrate [64]. This effect has to be distinguished from the formation of so-called breath figures which are formed at high relative humidity (over 60 %) due to water condensation on the evaporatively cooled polymer solution [65,66]. The breath figure technique can be applied to generate nearly hexagonally arranged arrays of holes [66] or for the fabrication of 3D structures [67]. Water droplets are introduced into the polymer solution film and leave behind holes after the film is solidified. These breath figure structures can be found both in films of one-polymer systems like PMMA in THF and of polymer blend systems.

In this article we present a method to obtain a polymer blend film with a purely lateral phase morphology, which means the blend separates completely into two lateral phases. The introduction of a small amount of water during the spin coating process is crucial for obtaining this purely lateral morphology. Either of the two different polymers can be dissolved independently afterwards by using a selective solvent. The remaining morphology is later on applied directly as a lithographic mask to fabricate nanopatterned self-assembled-monolayer (SAM) templates. Performed at higher humidity, our technique combines polymer blend phase separation with the breath figure formation. A three phase lithographic mask is formed in one process step, giving the chance to produce a SAM template with three different chemical functionalities.

### Polymer Blend Lithography

The polymer blend lithography method is demonstrated schematically in figure 1. The most important prerequisite is to have a polymer film consisting of two immiscible phases, which are laterally separated on the substrate. Here the polymer blend solution is prepared with the PS and PMMA dissolved in methyl-ethyl-ketone (MEK). As schematically shown in figure 1, it is found that this system decays into a purely lateral phase morphology during spin-casting of the solution at a moderate humidity, which means that both phases reach from the free air interface down to the silicon oxide substrate. This is by far not the common case. In most cases of polymer blend solution a mixture of lateral structures and a vertical phase morphology is formed. The result is also found for the PS+PMMA blend in MEK, if spin-casted in a dry atmosphere. Immiscibility qualifies for the possibility to dissolve selectively one component, which is on one hand important, if the other component is desired to be used as a lift-off mask. The miscibility on the other hand has the consequence that one component has a higher affinity to the substrate (hydrophilic) than the other one which prefers the free air interface (hydrophobic). The resulting morphology is a layered situation where the hydrophilic polymer wets the substrate while the hydrophobic most likely wets the free polymer-air interface. The upper layer becomes unstable and dewets so that droplets are formed. Therefore the final morphology is usually one phase "floating" in a lake of the other one. After the selective dissolution of the "floating" phase there is still a thin film of the other polymer in every hole, which is not the desired situation for polymer blend lithography.

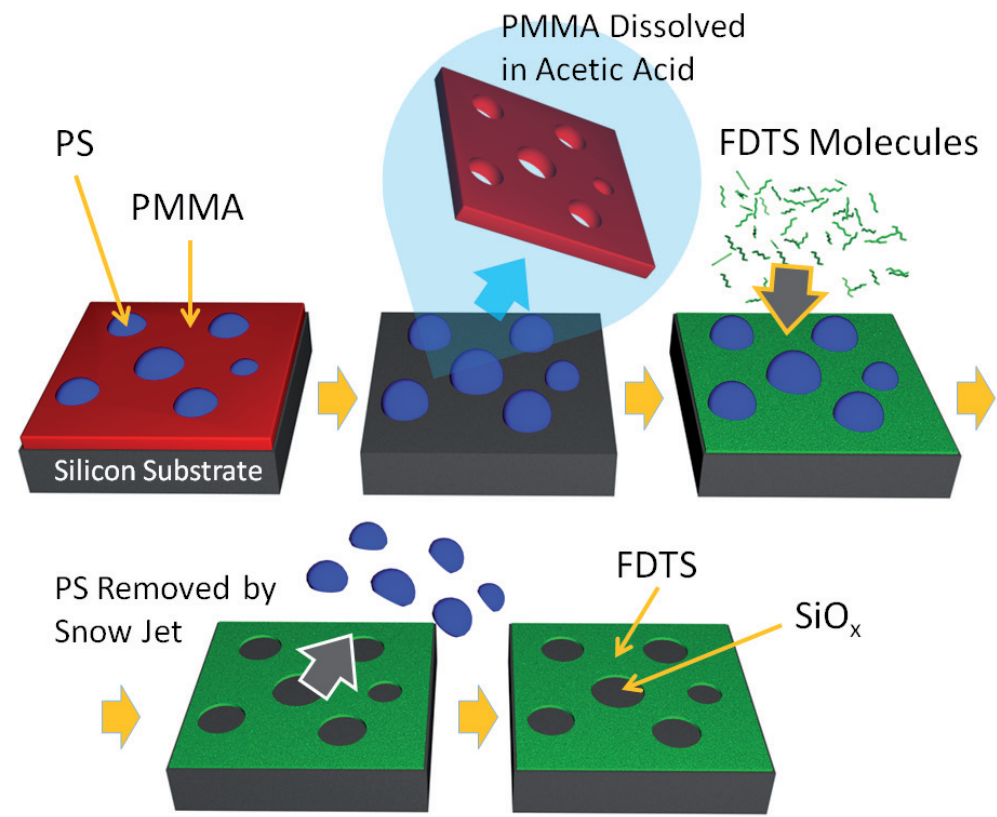

Fig. 1: Schematic drawing of the Polymer Blend Lithography process. After spin-coating in a controlled atmosphere, a purely lateral morphology of PS droplets (blue) in a PMMA matrix (red) is formed. After the dissolution of PMMA in acetic acid, the PS droplets remain and can be used as a mask for the deposition of a fluorine-terminated SAM (FDTS/green). Via a snow jet treatment the PS droplets are selectively removed and a patterned SAM is formed.

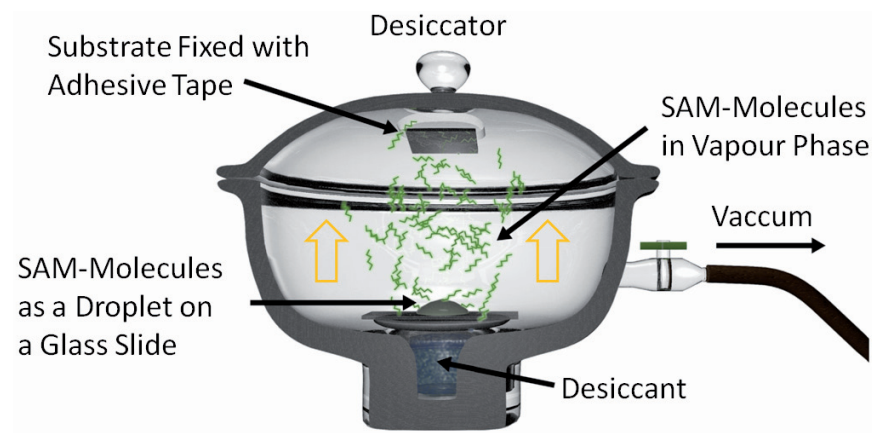

Fig. 2: Preparation of a densely packed SAM, performed in the vapor phase within a desiccator.

Here we present a recipe of how to create a purely lateral morphology without this drawback. The morphogenesis of this structure will be in the focus of a forthcoming publication. With the structure generated using the given recipe it is possible to remove one component (e.g. PMMA) and to deposit a SAM on the completely freed silicon oxide substrate areas with very high reproducibility. After the silane molecules have bonded covalently, the remaining polymer phase (PS) is removed. The deposition of the SAM is performed by vapor phase deposition [59] in a vacuum desiccator ( figure 2). During deposition, the samples are mounted face down on the lid of the desiccator. After the SAM is formed, the sample is removed from the vessel and the remaining polymer is removed via snow jet treatment. Consequently a "monolayer copy" of the original phase morphology is left with a topographic contrast of the height of the SAM, usually in the range of 1–2 nm, depending on the type of molecules used. By the choice of the SAM molecules the desired chemical surface functionality (functional group) can be defined.

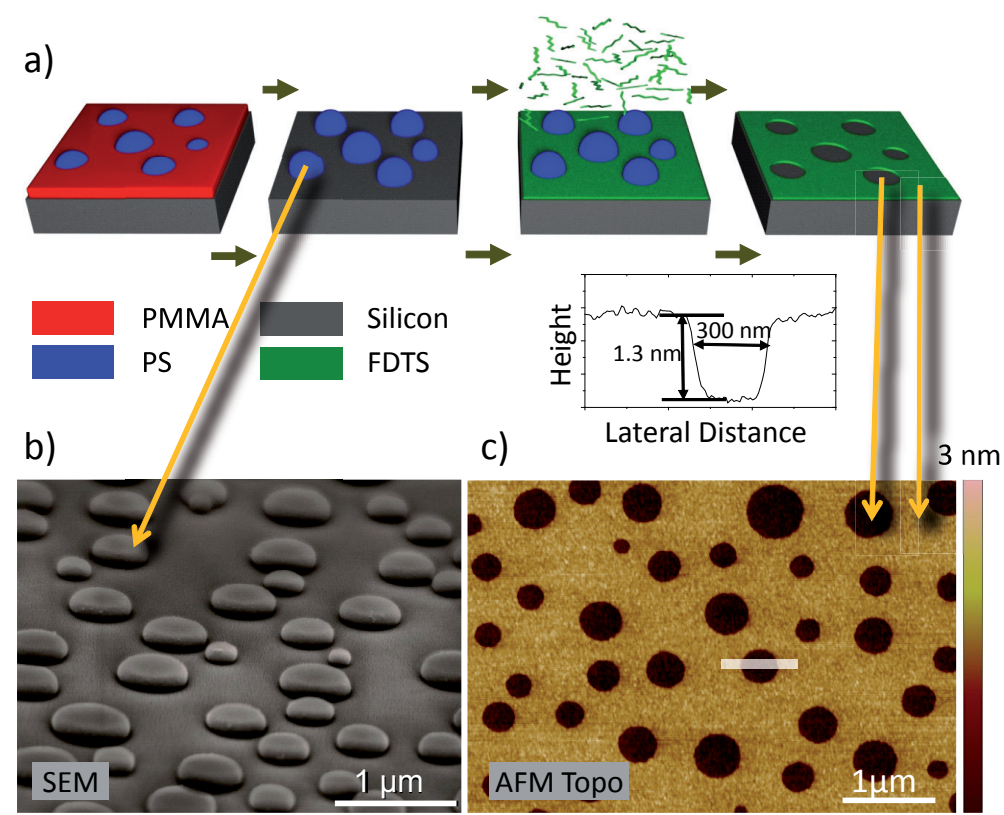

Fig. 3: Fabrication of a two-phase SAM template spun-cast at the humidity of 45%. (a) Schematic drawing of the process, silicon substrate (grey), PMMA (red), PS (blue) and FDTS (green). (b) SEM image of a polymer blend mask rinsed with acetic acid. (c) AFM image (retrace image measured in contact mode in liquid) of a two-phase SAM template. The cross section demonstrated is the average of the trace and the retrace images. The depth of the holes is 1.3 nm independent of the intensity and the duration of the snow jet treatment.

**Two-Phase Templates**

Via a spin coating process of a polymer blend solution at a humidity of 45 %, a purely lateral phase separated film consisting of the two polymer components is produced ( figure 3a). In  figure 3b an SEM image of a polymer blend mask rinsed in acetic acid is shown (the image was taken with a tilted angle of around 45°). After this treatment only the PS islands remain on the silicon substrate. The PMMA layer (marked red in figure 3a) has been completely removed. After the deposition of the 1H,1H,2H,2H-perfluorodecyl trichlorosilane (FDTS) SAM, the polymer islands were removed by a snow jet treatment. In figure 3c an AFM topography image of the remaining FDTS SAM template is shown. Each PS island leaves behind a hole in the mono-molecular layer. The average diameter of these holes is about 400 nm. The film has a topography of 1.3 nm. The depth of the holes is independent of the intensity and the duration of the snow jet treatment (see also suppl. info.) This indicates that the FDTS monolayer is well bound to the substrate and that the lift-off of the PS islands is complete.

**Island Size Tailoring**

The dependence of the PS island diameter upon the polymerization degree of PS is shown in figure 4. It can clearly be seen that the average diameter and the width of the diameter-distribution decrease with reducing molar mass of the polymer. When using PS of 9.58 kg/mol, the average diameter of the islands is about 90 nm and a very narrow diameter distribution from about 50 to 150 nm is obtained. For PS of 248 kg/mol an average diameter of about 500 nm and a wider diameter distribution from about 200 nm to 800 nm is found. Higher molar mass of the polymer increases the viscosity of the solution and consequently increases the film thickness and at the same time the height of the PS islands. All of these islands are formed during the spin-coating process in less than two seconds. Film drying kinetics is measured by an in-situ reflectometry technique performed with our laser setup described elsewhere [41]. Increased film thickness leads to a longer drying time, a larger domain size and a higher PS domain height, as clearly seen in figure 4b. This result shows that the molecular weight can be used as a parameter to adjust the domain size in the polymer blend lithography method. Besides the main structure size which can be reliably controlled there are always some small structures observed. In the histograms shown in Fig. 4c there is a detectable tail down to 90 nm for all molecular weights. This tail is most probably a signature of a secondary phase separation during the complex structure formation process.

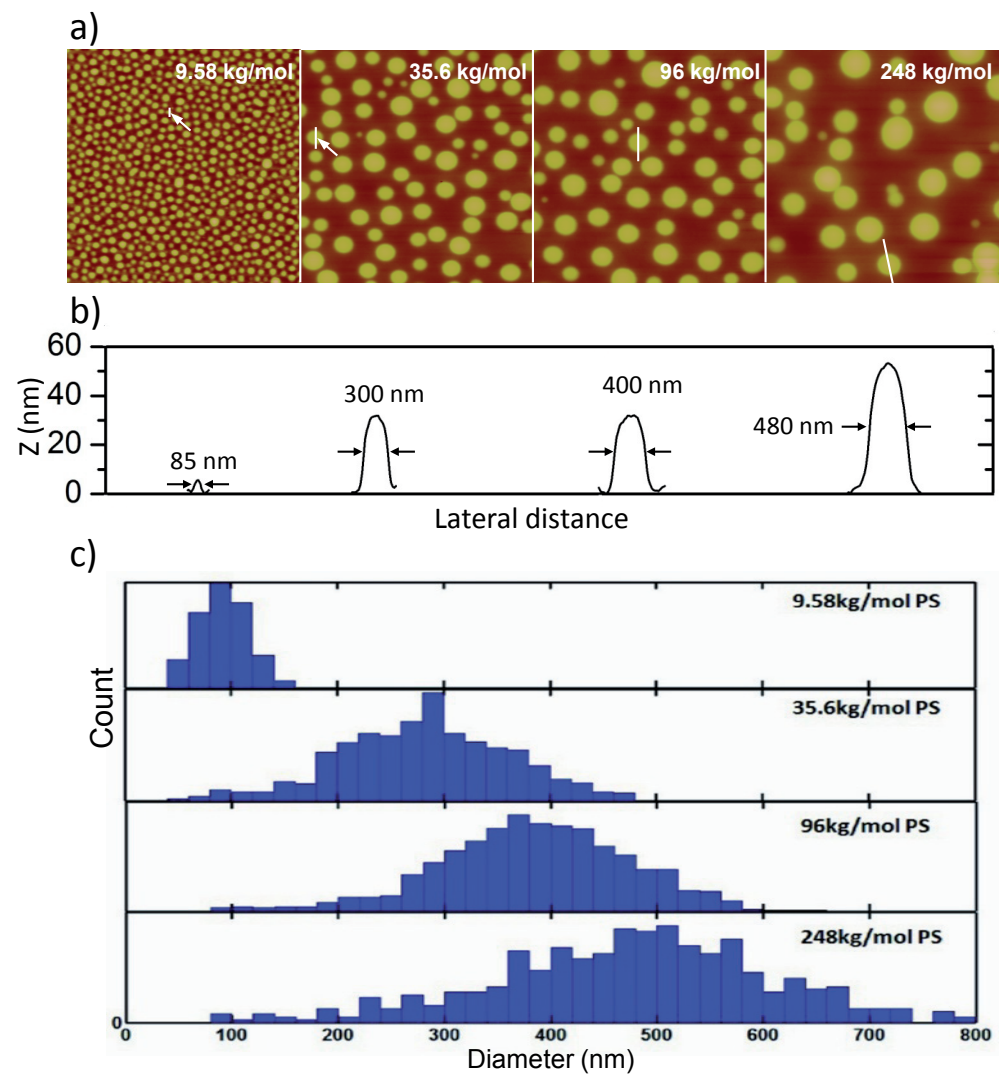

Fig. 4: Dependence of the PS island diameter and height by varying the molar mass of PS. (a) AFM images of a polymer blend film formed from various PS molar mass of 9.58, 35.6, 96 and 248 kg/mol. The scan areas of all AFM images are 5 × 5 µm². (b) Height profiles of selected PS islands of average size (height above the PMMA matrix level). (c) Distribution of the diameters of PS islands of various molar masses.

## Three-Phase Templates

For a range of relative humidity from 50 to 65 %, the resulting phase morphology is different from the situation shown in figure 3 (45 % humidity). As can be seen in figure 5b, holes in the polymer film can be observed directly after spin coating. Besides these open holes, there are smaller depressions and embedded PS droplets visible at the surface.

Due to the rapid evaporation of the solvent during the spin coating process the sample-surface is cooled down. At this highly increased humidity the sample-surface reaches the dew point. The result is that water condensates and then forms droplets which leave holes in the polymer film after it is solidified. The small depressions are most likely relics of smaller water droplets which did not reach the silicon substrate. Hence, the result of the spin coating process is a perforated PMMA layer with embedded PS droplets. This provides the opportunity to design a three-phase pattern as described below.

The (water-) holes can directly be filled with a silane monolayer. Here we used the APTES molecule exposing an amino-functional group. After removing the PMMA layer with acetic acid, the $CF_3$-terminated FDTS SAM was deposited in vapor phase. Next, we removed Polystyrene by snow jet treatment as described before. The FDTS as well as the APTES SAMs withstand this cleaning procedure without any detectable change at their surface, as can be seen in figure 5c. The three-phase SAM template consisting of APTES, FDTS and silicon oxide pattern elements is fabricated with topography of approximately 1.3 nm. The roughness of 0.2 nm remaining in the $SiO_x$-regions is in the same range as the one of the original Si-wafer. The height of the APTES-SAM was found to be 0.7 nm, measured in contact mode AFM in liquid. So the APTES regions look like halve-filled holes (see figure 5c).

## Perspectives of Polymer Blend Lithography

These patterned two-phase or three-phase surfaces which show a high chemical contrast and at the same time an extremely flat topography, make them an ideal template or platform for constructive lithography [4], cell adhesion studies or the study of other template-induced phenomena. The FDTS SAM can be replaced with other silanes such as octadecyltrichlorosilane (OTS) or polyethylene glycol (PEG) silane for desired applications [68,69]. The bare silicon surface at the bottom of the holes can be functionalized with another silane for certain applications. For example, in our recent publication the holes, filled with amino-propyltriethoxy-silane (APTES) were used for the growth of ZnO layers [4] by chemical bath depostition. Structured and non-structured ZnO layers are used e.g. in gas sensor applications [70–72]. Silane-based follow-up reactions can be used to produce silane multilayers [73], which only grow in the predefined areas. This type of SAM-template has also potential applications for the selective growth of titanium oxide and for graphene on surfaces [74,75] or cell adhesion studies [69]. Here without any further treatment we have generated an amphiphilic surface, featuring at the same time both hydrophobic (FDTS) and hydrophilic (APTES or SiO$_x$) areas. The versatile and fast preparation technique makes this approach attractive for many applications of such ultra flat nanopatterned surfaces.

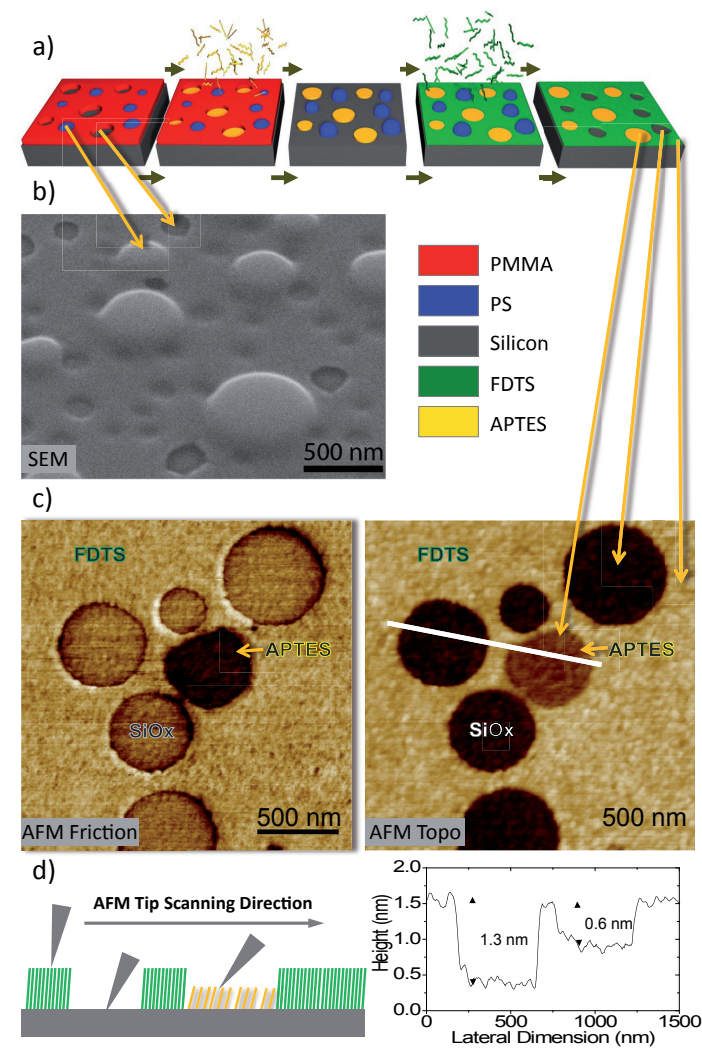

Fig. 5: Fabrication of a three-phase SAM template spun-cast at the humidity of 65%. (a) Schematic drawing of the process, silicon substrate (grey), PMMA (red), PS (blue), FDTS (green) and APTES (yellow). (b) SEM image of a polymer blend mask with breath figures. (c) AFM images (both retrace images) of a three-phase SAM template. The cross section shown here is the average of the trace and the retrace images. (d) Schematic drawing of the idea of an AFM friction image. The first SAM, which was deposited is APTES. Its height is half the height of the FDTS-SAM which was complemented after the PMMA mask has been removed. Finally, after the removal of the PS islands the remaining holes have a depth of 1.3 nm, which is independent of the intensity and duration of the snow jet treatment.

## 2.2   Materials and Methods

### 2.2.1   Patterned Substrates Polymer Blend Lithography

**Polymer Solution**

Poly(methyl methacrylate) (PMMA) (Mw=9.59 kg/mol PDI=1.05) and poly(styrene) (PS) (Mw=96 kg/mol PDI=1.04) were purchased from Polymer Standards Service GmbH (PSS) and dissolved directly in methyl-ethyl-ketone (MEK)(Aldrich). The total concentration of the two polymers was 15 mg/ml and the mass ratio between PS and PMMA was 3:7. For the demonstration of tuning the diameter of PS islands a set of polymer solutions were made with various PS molar masses: 9.58 kg/mol, 35.6 kg/mol, 96 kg/mol and 248 kg/mol. All other parameters were kept constant.

**Cleaning of Si Substrates and SAM Templates**

Silicon substrates were used as delivered with their native oxide layer. The substrates and the SAM templates were cleaned by the snow jet method [76]: the wafers were exposed to a jet of CO$_2$ ice crystals, which were produced by expanding CO$_2$ through a nozzle. In this way, surface contaminants are removed either by mechanical impact or by dissolution in CO$_2$.

**Preparation of a Polymer Blend Lithographic Mask**

The polymer blend films were spin-casted at a speed of ca. 1500 revolutions per minute (rpm) onto silicon substrates cleaned by snow jet treatment (at least 20 seconds for a 2 cm x 2 cm substrate). For the two-phase SAM templates, the relative humidity was set to 45 % during the spin-coating process and for the three-phase templates to 65 %. The humidity was controlled by venting the chamber (about 1 liter volume) with a mixture of nitrogen and water-saturated nitrogen (total flow rate approximately 40 standard cubic centimeters per minute (sccm)). The humidity in the chamber is measured by a hygrometer (Testo 635).

**Fabrication of SAM Templates**

For the two-phase template the PMMA was selectively dissolved by acetic acid, as shown in figure 1a and b. Samples were rinsed in the acid and constantly moved for 30 seconds. The samples were then rinsed two times and dried in a stream of nitrogen. The silane SAM was deposited over night in a desiccator containing two droplets of 1H,1H,2H,2H-perfluorodecyl trichlorosilane (FDTS) (Aldrich) and evacuated to a pressure of 50 mbar. The PS islands were later removed by snow jet lift-off. For sufficient impact it is important that the CO$_2$ gas cylinder is at room temperature and has a proper filling level. The polymer mask can be alternatively dissolved in THF, following the protocol described above for acetic acid.

For the three-phase template the amino-propyltriethoxy-silane (APTES) (Aldrich) SAM was deposited onto the silicon surface inside the holes of the lithographic mask in gas phase, shown in figure 5a and 5b. The PMMA was removed by acetic acid and the freed silicon surface was covered then by different silane molecule, FDTS with the same deposition method as APTES. The PS islands were removed by snow jet treatment as well. Instead of using a snow jet, the polymer mask can also be dissolved by organic solvent (tetrahydrofurane (THF)).

## Sample Characterization

The polymer blend masks were characterized by atomic force microscopy (AFM) and scanning electron microscopy (SEM). The AFM images were made with a commercial multimode system (DI Multimode IIIa) in tapping mode. The samples were scanned at ambient conditions immediately after removing them from the solution. SEM images were taken at 2 kV with a LEO 1530 SEM using a secondary electron detector. All AFM images of the SAM templates were taken in contact mode in the liquid cell filled with demineralized water (Bruker Dimension Icon-PT).

### 2.2.2 Homogeneous Substrates for Studies of Cell Substrate Interaction

For cell adhesion studies different homogenous substrates were prepared using thermal evaporation of metals, by spin casting polymers or by using surfaces of bulk materials.

Aluminum, gold and chromium were vaporized, in order to obtain a 20–30 nm thick layer. Regarding the samples with aluminium and gold, we also deposited 2 nm of chromium between the glass and these metals, in order to increase the adhesion to the glass. Fibronectin and collagen I were prepared in a density of about 2–5 μg/cm², using them before they dry out, otherwise their structure can change. They are in a liquid form, but dense enough to remain on the surface. We used PMMA with 25000 as molecular weight. Since the PMMA don't adhere very well on the glass, we first put a clean glass in the plasma cleaner and, after the spin coating of PMMA, we exposed the sample to UV light in order to break some chains: in this way the PMMA can be better adsorbed by the glass. The sample called plexiglass is referred to PMMA with mixed molecular weights.

The topography of these surfaces was measured by AFM. Figure 6 shows selected scans with sides of 20 μm. The surfaces show topographic structures with a variety of length scales. These structures enhance or dampen the chemical differences of these surfaces. For instance it changes the fractal dimension of the surface. Nevertheless, the roughness is extremely low (less than 30 nm corrugation), so that those surfaces can be considered flat and homogeneous when analyzing the behaviour of a living cell.

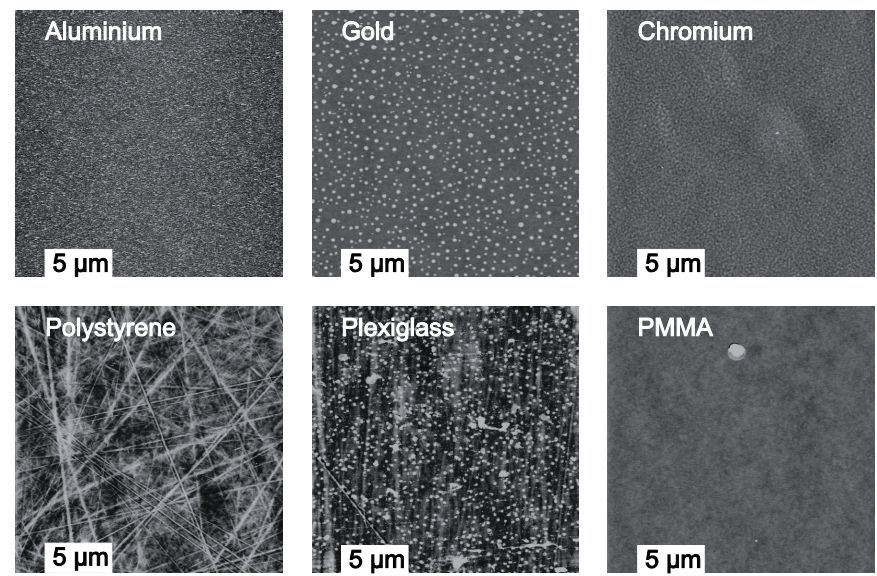

Fig. 6: a) Substrates for cell migration experiments. The AFM images are height coded. Black is low, white is high. The white bar represents a length scale of 5 μm. The height ranges from black to white are: 20 nm for aluminum, gold, chromium, polystyrene, 30 nm for plexiglass and 10 nm for PMMA.

The substrates were washed with ethanol and sterilized before using them in the cell culture.

### 2.2.3 Cell Culture

The so-called Panc-1 cell line (ECACC 87092802) is a human pancreatic carcinoma, epithelial-like cell line. For both the experiments regarding cell migration (scratch assay) and AFM measurements of their stiffness, Panc-1 cells were cultivated in standard DMEM (Dulbecco's Modified Eagle Medium) medium with 10 % Fetal Bovine Serum (FBS) at 37°C and 5 % $CO_2$ for at least 48 hours.

The cells were cultivated in petri dishes, directly on the bottom or on a cover glass where the substrate was prepared. For AFM measurements of cells cultivated on glass, collagen I and fibronectin we used chamber slides. The cell density for the AFM measurements was about $5 \times 10^4$ cells/ml on the day we splitted them. After an incubation time of two days we carried out the measurements. For the scratch assay we needed a confluent density of the cells on the surface. SK8/18 and SW-13 cells were cultivated on a glass covered with a monolayer of fibronectin, in order to increase the adhesion on the surface.

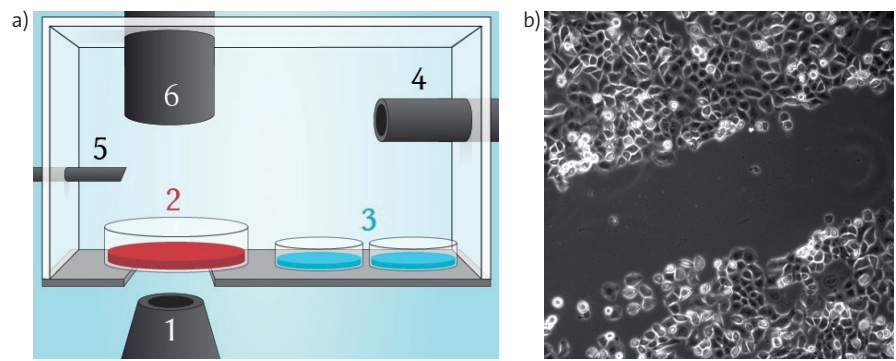

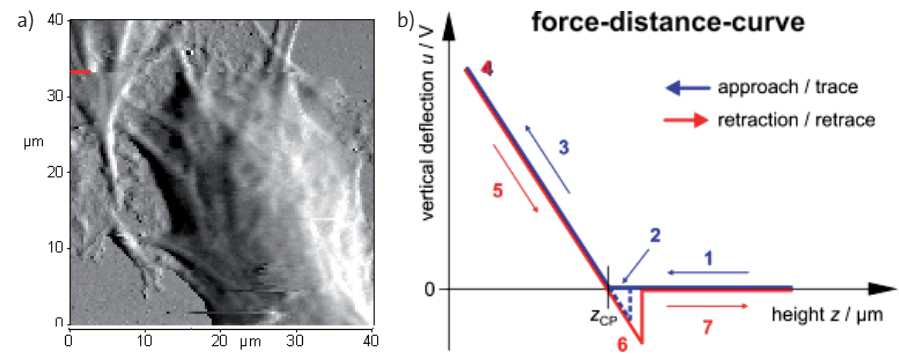

Fig. 7: a) Scheme of the environmental control chamber mounted on a Leica phase contrast microscope. (1) Objective (10×); (2) Petri dish containing our sample (the red fluid represents the medium for cell culture); (3) Water filled Petri-dishes; (4)/(5) $CO_2$ and temperature control; (6) Light source with shutter. b) Scratch assay of Panc1 cells cultivated on Chromium.

Fig. 8: a) AFM scan of SK8/18 cell cultivated on fibronectin. Even if a tip with a 2 µm sphere was used in this measurement, we can see the underlying structure in the upper part of the cell. b) Scheme of a force-distance curve on a hard sample.

### 2.3    Mechanical Properties of Cells on Different Substrates

#### 2.3.1    Atomic Force Microscopy on Living Cells

AFM [77] was used to image cells in physiological conditions, with minimal pretreatment and, potentially, with nanometer resolution. The microscope has a field of view of typically $100 \times 100$ µm in X and Y directions and about 15 µm in height [78]. The first AFM image of a living cell was published in 1990 [79]. Typical forces of 10–30 nN exerted by the cantilever tip on the cell membrane during the scan in contact mode do not cause excessive damage to the cell. Due to the softness of the cell membrane, the cantilever tip often follows the shape of the rigid cytoskeleton, and the underlying structure can be visible (see [5] and Figure 8 a)) [80–84] If dimensions, geometry and spring constant of the tip are known, it is possible to estimate the stiffness of a cell and also the forces both within and between. In this way measurements of mechanical properties and physical interactions can be done for different materials and biological samples from the nanometer to the micrometer scale.

Figure 8b shows a schematic force-distance curve related to a hard surface. At point 1 the probe is far away from the sample, there are no surface forces and the deflection is zero. While the probe is getting close to the surface (point 2), the cantilever usually starts to deflect due to attractive Van der Waals forces. When the gradient of the attractive force exceeds the spring constant of the cantilever, the probe jumps in contact with the surface. From this point, the deflection of the cantilever will increase with the increasing applied force (point 3), either in a linear or not linear way in case of hard or soft samples respectively. When the maximum force applied to the sample, which is determined by the user, is reached (point 4), the cantilever is retracted, moving again in parallel with the sample. If there are adhesive forces between probe and sample, the cantilever will bend downward (point 6) until the restoring force of the cantilever exceeds the adhesion force and the cantilever snaps back to its equilibrium position (point 7).

Our purpose was to study the behavior of Panc-1 cells on different surfaces and to compare it with the cell lines sw-13 and sk-8/18. Measurements of the cell stiffness and a detailed analysis of the cell migration were carried out in order to find some parameters that can clearly distinguish the different cell-substrate interactions. In order to check if different substrates can induce some changes in the internal structure of the cells, we measured their cortical stiffness with an AFM.

Tab. 1: Young's moduli of the cells measured on the center of the cell and in the middle between the center and the rim. Different values for the pancreatic cells on different surfaces has been observed.

| | Intermediate point [kPa] | Center [kPa] |
|---|---|---|
| Plexiglas | 3.13 ± 0.37 | 5.54 ± 0.52 |
| PMMA | 2.51 ± 0.43 | 6.48 ± 2.45 |
| Collagen I | 2.48 ± 0.28 | 3.72 ± 0.62 |
| Chromium | 2.19 ± 0.32 | 4.77 ± 1.3 |
| Polystyrene | 1.59 ± 0.15 | 3.67 ± 0.96 |
| Glass | 1.48 ± 0.29 | 3.28 ± 1.25 |
| Aluminium | 1.44 ± 0.24 | 2.80 ± 0.45 |
| Fibronectin | 1.33 ± 0.29 | 3.19 ± 1.17 |
| Gold | 1.31 ± 0.24 | 2.72 ± 0.31 |
| sk-8/18 | 2.14 ± 0.55 | 3.35 ± 0.60 |
| sw-13 | 1.44 ± 0.30 | 2.62 ± 0.42 |

For each sample, 8–10 cells were measured, scanning at first the whole cell, applying a force of about 4–5 nN, then saving the Force-Distance Curves (FDC) on three different points: one on the center of the cell, one on the rim and one halfway between the center and the rim (the so-called "intermediate point"). On each point, about 15 curves were saved, in order to obtain a more reliable value.

Regarding the FDCs, the indentation for every measurement was 1–1.5 µm and the maximum force applied was 3–4 nN. Since the height of the measured cells was about 4–5 µm, we did not see the influence of the substrate in the FDCs, except of course for the points on the rim of the cells. Therefore the Hertz model can be used to estimate the Young's moduli of our samples, which predicts the following relation:

$$F = \frac{4\sqrt{R}}{3\pi}\left(\frac{E}{1-\nu^2}\right)\delta^{3/2}$$

where F is the force applied on the sample, R is the radius of the probe, E

is the Young's modulus, ν is the Poisson's ratio of the material (in our case equal to 0.4) and δ is the indentation into the surface of the sample. The data obtained from these measurements are showed in table 1.

We can observe different values of the stiffness for Panc-1 cells on the different substrates, thus we have found a property that emphasize the different cell-substrate interactions. Regarding the cell lines sw-13 and sk-8/18, a different stiffness has been calculated for each cell type, even if there is not a clear difference between their stiffness and the values regarding Panc-1 cells.

These results seem to confirm the hypothesis that the cell tries to adapt itself to the environment by remodeling its cytoskeleton. Pancreatic cells seems to be stiffer on the two PMMA samples with different molecular weights and on collagen I, and softer on aluminium, fibronectin and gold.

Since sw-13 and sk-8/18 cells have less intermediate filaments than pancreatic cells, we expected that they were softer than Panc-1. However, considering the error bars, we can not really distinguish between Panc-1 cells and the other two cell lines with this parameter. A possible explanation could be that they react to the same substrate in a different way. Nevertheless, we can observe that sk-8/18 are stiffer than sw-13, due to the fact that sw-13 are basically the same kind of cells as sk-8/18 but without keratin 8/18.

We favour the hypothesis, that the extracted cytoskeleton lacks the stabilization effect of the cell membrane. The AFM measurement not only probes the cytoskeleton and its response, but also the viscoelastic response of the cytosol enclosed in the cell membrane. This stabilizing effect increases the overall Youngs modulus. Microrheology only probes the cytoskeleton without the membrane.

### 2.3.2 Microrheology of Extracted Keratin Networks of Panc-1 Cells for Mechanical Testing

We complemented the measurements of the mechanical stiffness of the cells with microrheological measurements [85]. Polystyrene beads with a diameter of 1 µm (Cat No. 4009A , Thermo Fisher Scientific, Waltham, MA, USA) were added to the cell culture medium. A home built microscope equipped with an oil immersion objective (CFI Apo TIRF 100XH, numerical aperture 1.49, Nikon, Tokyo, Japan) and a high-speed camera with a frame rate of 5000 Hz (MotionX Pro 4, Imaging Solutions, Regensdorf, Switzerland) recorded the Brownian motion of the microspheres in the extracted keratin networks[86][87]. Particle displacements as small as 7 nm were detected. Multiple particles were tracked at the same time within the same image to minimize the impact of noise on the variability of the measurement within the sample.

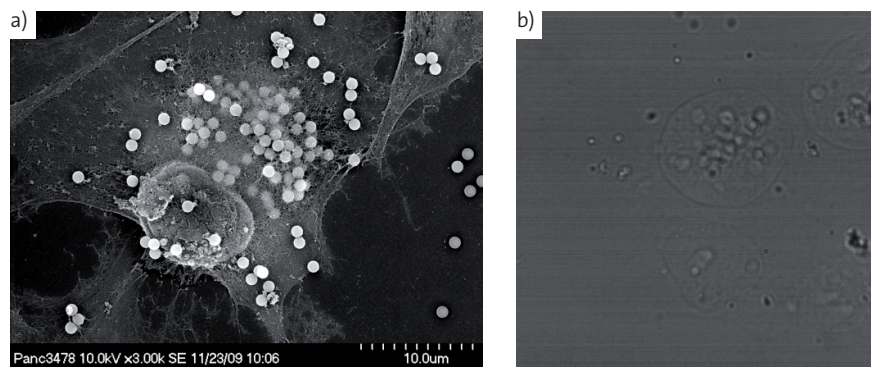

Fig. 9: a) Extracted keratin cytoskeleton of PANC-1-cells with 1 µm polystyrene beads.
b) optical micrograph of 1 µm polystyrene beads in extracted keratin cytoskeletons.

Figure 9 a) shows an overview of the extracted keratin network of a Panc-1 cell. This network. The nucleus is clearly visible. Apart from the nucleus the network is mostly free of cytoskeleton molecules [87]. The keratin network is almost free of contrast in the optical microscope. Therefore the diffraction pattern (an Airy-disk) of the beads is virtually not affected by the substrate (see figure 9b). The mean square displacement of the Brownian motion of each bead is calculated. Via a Laplace-transform the complex elasto-mechanical moduli can be calculated. Figure 10 shows the measurement of G′ and G″ in the cytoskeleton of extracted pancreatic cancer cells (PANC-1 cell line). Real (G′) and imaginary (G″) parts are linked by the Kramers-Kronig-relation, a consequence of causality. The storage modulus, the spring like component, is rater frequency independent. The loss modulus on the other hand shows characteristic peaks which are linked to characteristic time constants in the network. The AFM in contrast measures the moduli at zero frequency The analysis of the data gave G ≈ 0.1 Pa. This value has to be converted to elastic moduli E. Assuming isotropy the conversion factor is calculated with a poisson ratio of 0.5. The values are much lower than those extracted from the AFM data. This reflects the missing of the cell membrane and of most of the cytosol.

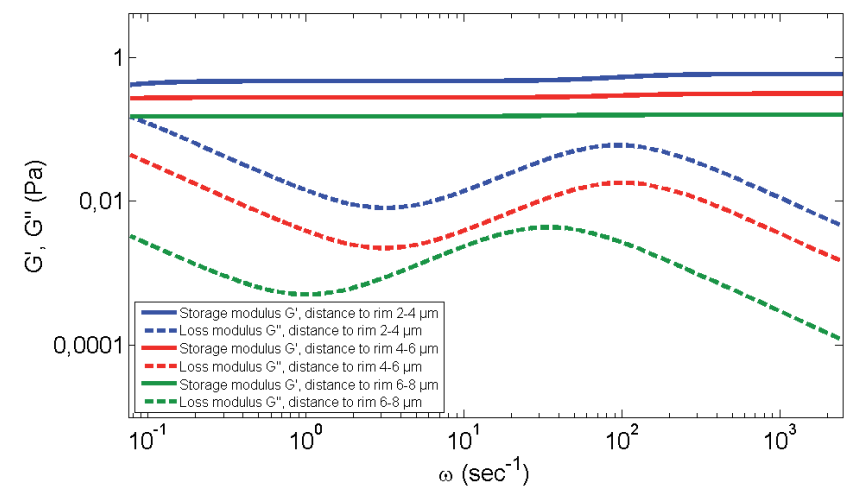

Fig. 10: Microrheological measurement oft he frequency dependent shear moduli (storage (G′) and the loss modulus (G″)) measured in the extracted cytoskeleton of Panc-1 cells. The two groups of three curves demonstrate the variation of the mechanical properties with the distance to the nucleus(near the nucleus (blue), at intermediate distances (red) and towards the rim of the cell (green)).

## 2.4    Cell Migration Experiments

The measurements of cell migration using the scratch-wound assay has been carried out and analyzed in detail. The movements of 20−25 cells were tracked for each sample, saving their position for each step, namely every 12 minutes, for one day measurement. As the positions of the cells are known as a function of time, their average velocity can be calculated between each step. From these information, velocity distributions, average velocities and autocorrelation functions can be evaluated.

### 2.4.1  Velocity Autocorrelation Function

The velocity autocorrelation function $\upsilon_{ac}$ (t) describe the self-similarity of the

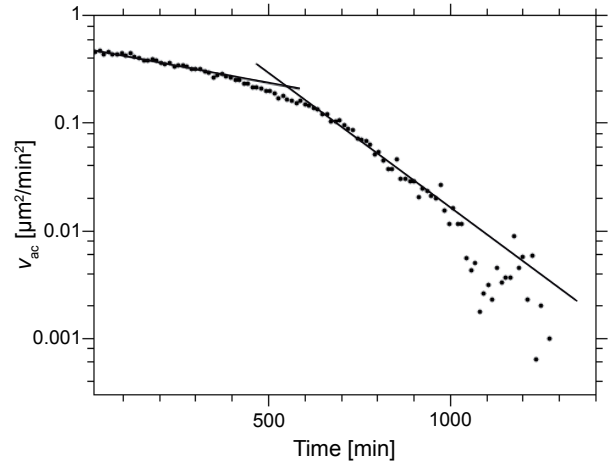

Fig. 11: Fit of the velocity autocorrelation function for Panc-1 on gold. The fitting line at short times is related to the first exponential decay and the line at longer times is related to the fit of the second exponential decay. Each regime can be fitted with a simple exponential decay.

velocities between two different times t and $t_0$ :

$$\upsilon_{ac}(t) = \frac{<\upsilon(t+t_0)\upsilon(t_0)>}{<\upsilon^2(t_0)>}$$

It is worth to notice that it was necessary to subdivide the tracked cells on the PMMA substrate into three groups, due to their significantly different behavior: the group a include 13 cells, group c 9 cells, group b only 2, and this explain why we will not show the values of the latter group in the following analysis. This means that the different values are due to a different behavior of the same kind of cells on the same substrate. No apparent difference could be observed on the video of cell migration between those cells. This can be related to the healthy state of the cell, to different cell cycles, or to some local dishomogeneity of the substrate morphology.

Figure 11 shows the velocity correlation function for Panc-1 cells on gold. There are two different regimes: each regime can be fitted with a simple exponential decay. Other functions such as inverse power law, stretched exponential and the sum of two exponentials can not fit the data as good as the function we choose to use.

Tab. 2: Values obtained from the fit of the first exponential decay of the velocity autocorrelation functions and the time between the first and second decay ($t_{break}$).

| | $t_{break}$ [min] | $A_1$ [µm²/min²] | $P_1$ [min] |
|---|---|---|---|
| Glass | 700 | $0.504 \pm 3.37 \cdot 10^{-3}$ | $662.25 \pm 9.74$ |
| Collagen I | 550 | $0.594 \pm 5.09 \cdot 10^{-3}$ | $900.9 \pm 27.19$ |
| Plexiglas | 900 | $0.474 \pm 3.16 \cdot 10^{-3}$ | $909.09 \pm 13.64$ |
| Chromium | 600 | $0.418 \pm 3.5 \cdot 10^{-3}$ | $990.1 \pm 27.84$ |
| Gold | 450 | $0.499 \pm 5.26 \cdot 10^{-3}$ | $561.8 \pm 15.65$ |
| PMMA group a | 800 | $0.498 \pm 4.4 \cdot 10^{-3}$ | $900.9 \pm 18.59$ |
| PMMA group c | 650 | $0.502 \pm 9.23 \cdot 10^{-3}$ | $462.96 \pm 13.67$ |
| Aluminium | 650 | $0.495 \pm 3.84 \cdot 10^{-3}$ | $813.01 \pm 18.11$ |
| Fibronectin | 650 | $0.538 \pm 5.15 \cdot 10^{-3}$ | $709.22 \pm 15.34$ |
| Polystyrene | 600 | $0.545 \pm 5.33 \cdot 10^{-3}$ | $440.53 \pm 7.04$ |
| sk-8/18 | 1000 | $0.247 \pm 5.25 \cdot 10^{-3}$ | $512.82 \pm 18.65$ |
| sw-13 | 900 | $0.255 \pm 6.31 \cdot 10^{-3}$ | $373.13 \pm 15.32$ |

The velocity correlation functions of those cells are characterized by exponential decays, and the values of the persistence times (see table 2) are different for each sample. The pancreatic cells on chromium, plexiglas, collagen I and the cells of group a on PMMA have the longest first persistence time, and this is reflected also in the persistence times of the second decays (data not showed). This means that Panc-1 cells on these surfaces tends to maintain the same velocity for longer times than on the other surfaces and also relative to cell types sk-8/18 and sw-13.

It is interesting to remark that there are two separated exponential decays for Panc-1 cells on all the substrates on which we cultivated them. This means that we have two different behaviors correlated in two different time scales. This could be related to the different cell cycles, or, if we consider the correlations between long times, they could have changed some of their characteristics in order to adapt to that environment. Maybe the correlations in the velocities for one time scale is related to the single cell motility and the correlations between the other time scale has something to do with the coordinated movement of the cell monolayer considered as one motile unit.

It is interesting to notice that sk-8/18 and sw-13 do not show up the second exponential decay (see figure 12). An explanation could involve the intermediate filaments, since this is the most remarkable structural difference

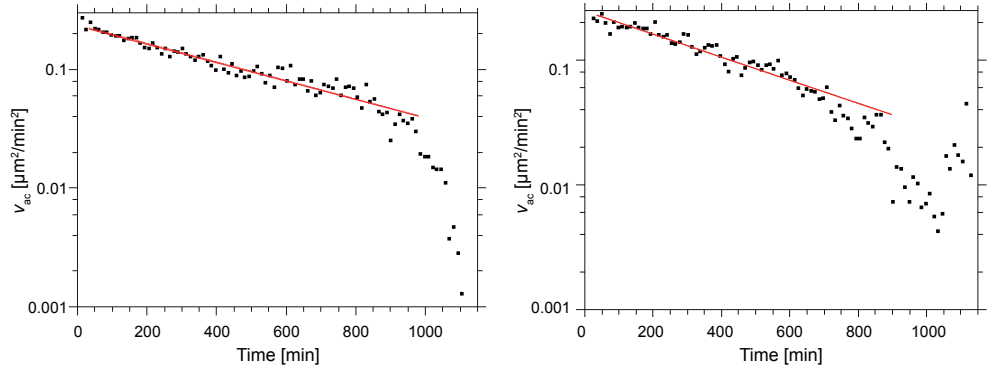

Fig. 12: Decay of the velocity autocorrelation function for sk-8/18 (left) and sw-13 (right). Those cell lines are characterized by velocity autocorrelation functions with only one exponential decay.

between those cell lines and the pancreatic cells.

Even if a whole class of models can provide velocity autocorrelation functions that are characterized by an exponential behavior, the Ornstein-Uhlenbeck (OU) process is the simplest and most commonly used for persistent random motion of motile cells. This process is characterized by the Fürth's formula

$$<\vec{d}(t)^2 >= 2nD(t - P(1 - e^{-t/P}))$$

for the mean square displacement $<\vec{d}(t) >=< \vec{r}(t) - \vec{r}(0)$. $< \dots >$ denotes expectation value, t is the time and n=1, 2 or 3 is the dimension of the space in which trajectories are studied. D can be defined as the motility coefficient for a micro-organism. P is the persistence time, and it represents the time for which a given velocity is "remembered" by the system [88]. From this formula, if we differentiate twice, we can obtain the velocity autocorrelation function

$$\phi(t) \equiv <\vec{v}(t) \cdot \vec{v}(0) >= \frac{nD}{P} e^{-|t|/P}$$

where $\vec{v} \equiv d\vec{r} / dt$.

### 2.4.2 Velocity Distribution on Different Substrates

The histograms of the velocities distributions can give us some more information about the underlying dynamics of the cell migration. The non-extensive entropy introduced by Tsallis [89,90]

$$s_q = \frac{1 - \int dx [f(x)]^q}{q - 1}$$

can give a generalization of Maxwell-Boltzmann thermodynamics. In the equation, f (x) represents the probability distribution function and q is a parameter that quantifies the degree of non-extensity. If this function is optimized under normalization and mean energy constraints, a generalized probability density function as well as the q-distribution of velocities [91] can be calculated:

$$F(\mathbf{v}) = A_q [1 - (1 - q) \frac{\alpha m \mathbf{v}^2}{2}]^{1/1-q}$$

Here v represents the velocity of the diffusing species and q is the entropic index. In this case $\alpha$ and m do not represent the actual temperature or mass, but they are related to an effective mobility. It is important to remark that the temperature is not that of the surrounding, because the motion is not thermally driven, but it can be considered as an effective temperature representing the cytoskeletal activity driven by nucleotide hydrolysis.

The limit of $q \to 1$ recovers the regular Maxwell-Boltzmann thermodynamics with Gaussian velocity distributions. We used the q-distribution fitting function in the form [92]

$$F(\upsilon) = \frac{a}{(1 + b\upsilon^2)^c}$$

obtaining the parameter q from the relation q = c + 1/c. Other functions such as simple exponential, power law, stretched exponential and Gaussian distribution can not fit the data as good as the q-distribution. Figure 13 shows the velocity distribution of Panc-1 cultivated on a fibronectin layer.

The velocity distributions can be fitted with a q-distribution. In order to verify what dynamics can describe the behavior of our cells, we need to find a relation between q and the exponent $\beta$ of the mean square displacement. Solving the Fokker-Planck equations represents another way to obtain probability distribution functions in the case of anomalous diffusion. A diffusion equation with fractional derivatives and a non-linear Fokker-Planck equation can describe the Levy-type diffusion [93][94] and the correlated anomalous diffusion [95–97]respectively.

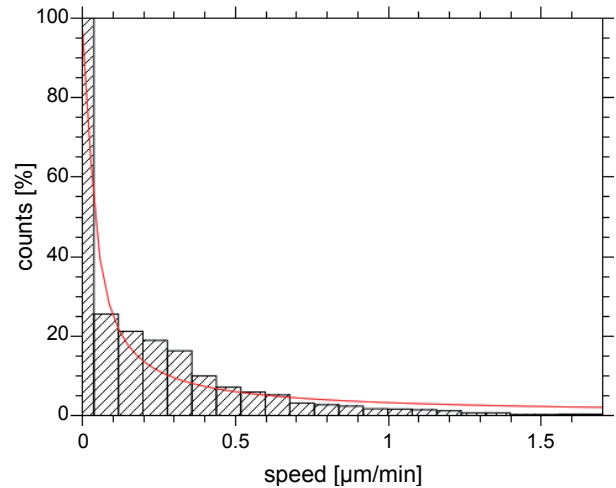

Fig. 13: Velocity distribution for Panc-1 on fibronectin fitted with a q- distribution function.

Tab. 3: Values obtained fitting the speeds distributions of all samples with the q-distribution

|  | $c$ | $q$ | $\beta = q-1$ |
|---|---|---|---|
| Aluminium | 0.522±0.055 | 2.44±0.25 | 1.44±0.25 |
| Collagen I | 0.404±0.042 | 2.88±0.30 | 1.88±0.30 |
| Chromium | 0.549±0.072 | 2.37±0.30 | 1.37±0.30 |
| Fibronectin | 0.463±0.039 | 2.62±0.22 | 1.62±0.22 |
| Glass | 0.472±0.046 | 2.59±0.25 | 1.59±0.25 |
| Gold | 0.557±0.048 | 2.35±0.20 | 1.35±0.20 |
| Plexiglas | 0.550±0.068 | 2.37±0.29 | 1.37±0.29 |
| PMMA group a | 0.556±0.065 | 2.35±0.27 | 1.35±0.27 |
| PMMA group c | 0.470±0.063 | 2.60±0.34 | 1.60±0.34 |
| Polystyrene | 0.498±0.049 | 2.51±0.24 | 1.51±0.24 |
| sk-8/18 | 0.565±0.053 | 2.33±0.21 | 1.33±0.21 |
| sw-13 | 0.570±0.038 | 2.32±0.15 | 1.32±0.15 |

After fitting the histograms, the parameter q was calculated for each sample and the results are summarized in Table 3.

In our case, only if we use the relation for two-dimensional Levy flights $\beta = q - 1$ [98] we obtain resonable values, comparable to the values obtained by fitting the mean square displacements, considering the experimental error. This means that the behavior of the tracked cells is characterized by a Levy- type diffusion, which can be described by a diffusion equation with fractional derivatives. This may be due to the interactions between cells, inducing correlations which modify the dynamics.

### 2.4.3 Average Velocity of Cells

The average cell velocity represents another important parameter that can distinguish all the different cell-substrate interactions analyzed, because it is different for each sample. Except for plexiglas and the group c of PMMA, the results for Panc-1 can be divided in two main groups. A first group, composed of polymers and biopolymers, on which the cells have a higher average velocity. A second group of inorganic materials, i.e. aluminium, glass, gold and chromium, which are characterized by a lower velocity of the cells. These results confirm the behavior observed in previous works [99.100], which found that cells exert less tension on softer gels but crawl faster comparing to stiffer substrates.

We can notice that the velocity of the cell lines sk-8/18 and sw-13 are smaller than the velocities of Panc-1. Even if these type of cells are different from Panc-1, this significant difference in the average velocity remark that maybe also the intermediate filaments play a significant role in cell motility.

### 2.4.4 Characteristic of the Distribution of Cell Positions (Kurtosis)

The kurtosis values of the distribution of the cells positions offer another possibility for determining whether the observed non-Gaussian statistics arise due to correlated or Levy-type anomalous diffusion. If the kurtosis equals the value 3, the distribution has a Gaussian shape. If it is smaller than 3, the distribution is less peaked than a Gaussian, and if it is bigger than 3 it is more peaked than a Gaussian distribution.

We calculated this parameter of the position distributions from the Mean Square Displacement (MSD) as it follows:

Fig. 14 : The kurtosis of the position distributions of all the samples analyzed is plotted as a function of time. It has been calculated from the MSD.

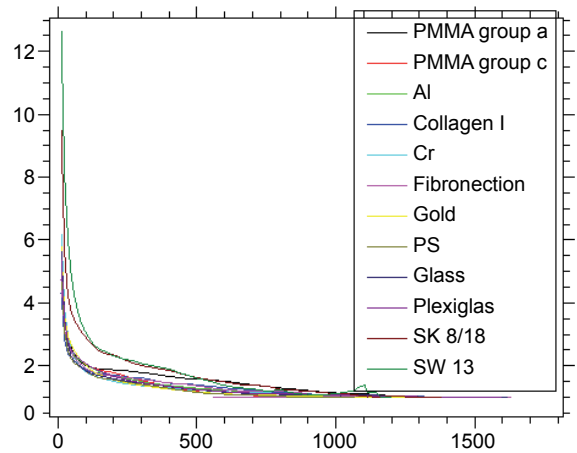

|  | $\beta$ |
| --- | --- |
| Glass | $1.500 \pm 0.003$ |
| Collagen I | $1.470 \pm 0.006$ |
| Plexiglas | $1.570 \pm 0.005$ |
| Chromium | $1.540 \pm 0.004$ |
| Gold | $1.540 \pm 0.006$ |
| PMMA group a | $1.490 \pm 0.007$ |
| PMMA group c | $1.490 \pm 0.002$ |
| Aluminium | $1.590 \pm 0.008$ |
| Fibronectin | $1.450 \pm 0.003$ |
| Polystyrene | $1.600 \pm 0.002$ |
| sk-8/18 | $1.180 \pm 0.004$ |
| sw-13 | $1.320 \pm 0.005$ |

Tab. 4: Average speed values for every sample. The error is the standard error of the mean.

|  | Mean velocity [µm/min] |
| --- | --- |
| Collagen I | $0.255 \pm 0.006$ |
| PMMA group a | $0.239 \pm 0.007$ |
| Polystyrene | $0.216 \pm 0.007$ |
| Fibronectin | $0.212 \pm 0.006$ |
| Aluminium | $0.207 \pm 0.005$ |
| Glass | $0.195 \pm 0.005$ |
| Gold | $0.183 \pm 0.005$ |
| Chromium | $0.182 \pm 0.005$ |
| Plexiglas | $0.182 \pm 0.004$ |
| PMMA group c | $0.135 \pm 0.007$ |
| sk-8/18 | $0.069 \pm 0.003$ |
| sw-13 | $0.059 \pm 0.003$ |

$$k(t) = \frac{\langle msd^2(t) \rangle}{\langle msd(t) \rangle^2}$$

In Figure 14 this parameter has been plotted for each sample as a function of time. We can see that the distributions of the cells positions tend to be more peaked than a Gaussian for short times but got rather broad (less peaked than a Gaussian) for long times, which is typical for super-diffusion processes. Moreover, the broad distribution of the displacements for long times confirms that the migration of the tracked cells follows a Levy flight dynamics.

### 2.4.5 Directionality of the Cell Movement

The features of anomalous diffusion include history dependence, long-range correlation and heavy tail characteristics for a random walk [101,102]. Mean square displacement (MSD) is an important tool to characterize anomalous diffusion processes [103–105], because it can quantify the directionality of the motion. Considering the following solutions of the Fokker-Planck equation [106]

$$msd \propto \begin{cases} (n\tau)^1 & \text{for a ballistic walk} \\ (n\tau)^2 \\ (n\tau)^\beta, 1 < \beta < 2 & \text{else} \end{cases}$$

where t=nτ , we can calculate the parameter $\beta$ for all our data sets (see table 5), fitting them with a power law function

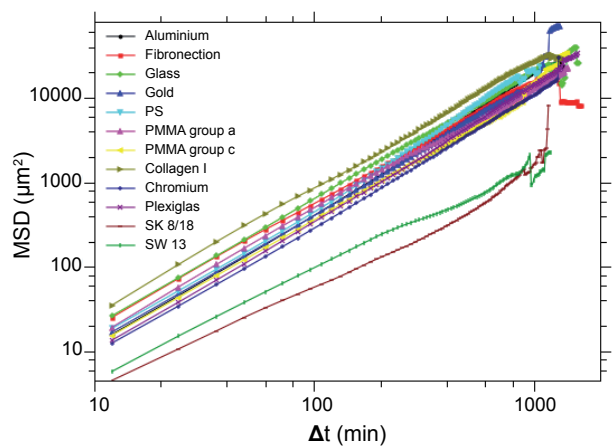

Fig. 15: Mean MSDs for Panc-1 on different substrates and for sk-8/18 and sw-13.

$$msd \sim t^{\beta}$$

The exponent $\beta$ in the power law of the MSD gives us the information about the directionality of the motion and thus it quantifies the randomness of the trajectories. Indeed, if $\beta < 1$ we have sub-diffusion: this means that we have a very high probability that the cell changes its direction from one step to the next. If $\beta = 1$ we have random motion, and if it is equal to 2 we have ballistic movement. For $1 < \beta < 2$ the movement is characterized by super-diffusion, and the higher is this value, the higher is the probability that the cell keeps the same direction relative to the last step.

The exponent $\beta$ of the MSD is not significantly different for the pancreatic cells cultivated on different surfaces, but they were clearly different from the values of the other two cell types. This means that this parameter can distinguish between different cell lines. Moreover, the calculated values of this parameter confirmed that the migration of the tracked cells is characterized by super-diffusion.

## 3  Conclusion

The purpose of this work was to study the behavior of Panc-1 cells on different surfaces and to compare it with the cell lines sw-13 and sk-8/18. Measurements of the cell stiffness and a detailed analysis of the cell migration were carried out in order to find some parameters that can clearly distinguish the different cell-substrate interactions.

With the AFM, the stiffness of the cells has been measured, and different values for the pancreatic cells on different surfaces has been observed. Regarding the cell lines sw-13 and sk-8/18, a different stiffness has been calculated for each cell type, even if there is not a clear difference between their stiffness and the values regarding Panc-1 cells. These mechanical data of the cells were complemented by our results of microrheology of extracted cytoskeletons on the mechanics of cytoskeletons.

The comparison of the migration data with the Young's modulus data show, that there is a vague connection between the velocity of the cells and their mechanics. The main influence, however, is the substrate.

One can speculate that the interaction with the substrate depends on the surface properties. These in turn will govern the number of adhesion points. They in turn affect the cytoskeleton. Hence our data support the hypothesis that the nature and the structure of the substrate can alter the cell migration velocity.

Then measurements of cell migration using the scratch-wound assay has been carried out and analyzed in detail. The velocity correlation functions of those cells are characterized by exponential decays, and the values of the persistence times are different for each sample. Interestingly, the velocity correlation functions of pancreatic cells are characterized by two separated exponential decays, unlike the other two cell lines that have only one exponential decay: this may regard the absence of intermediate filaments in sw-13 and sk-8/18 cells, which seems to play a not negligible role in the cell migration. The average cell velocity represents another important parameter that can distinguish all the different cell-substrate interactions analyzed, because it is different for each sample. Moreover, our data confirmed the results of previous works, in which a higher cell velocities on softer substrates was observed. The velocity distributions can be fitted with a q-distribution. Moreover, a further analysis brings to the conclusion that the observed cell migration can be described as a Levy-type diffusion. This has been confirmed also from the kurtosis of the distributions of cell positions. The exponent $\beta$ of the MSD is not significantly different for the pancreatic cells cultivated on different surfaces, but they were clearly different from the values of the other two cell types. This means that this parameter can distinguish between different cell lines. Moreover, the calculated values of this parameter confirmed that the migration of the tracked cells is characterized by super-diffusion.

*Polymer blend lithography* (PBL) makes use of the lateral structure formation during the spin coating process of a polymer blend film. The structures are transformed into a patterned SAM with two or three different chemical functionalities by a lift-off process. PBL starts with spin-casting a polymer blend (e.g. PS/PMMA in MEK) onto a substrate at a defined relative humidity. By selecting adequate conditions, a polymer blend film with purely lateral phase morphology is formed. After the selective dissolution of one of the polymer components, the remaining second polymer component can be directly used as a lithographic mask. This lithographic mask, in turn, can be removed by snow jet lift-off after deposition of a silane monolayer (SAM) on the unprotected areas in the vapor phase. For the examples demonstrated, the fabricated nanopatterned template shows a chemical contrast between the functional group of the silane SAM and the bare silicon oxide. This quasi two-dimensional pattern has about 1 nm topography. The bare silicon oxide surface can be filled with another silane SAM for specific applications. The lateral structure size within the nanoscale pattern is determined by the diameter of the PS islands formed during the spin coating process. The mean value of the statistically distributed diameters of PS islands can be varied between 90 nm and 500 nm by changing the molar mass of the PS moiety. Combined with breath figures, this lithographic method can even be used for the fabrication of three component templates. Here we use for the patterning the $CF_3$-terminated FDTS monolayer, the amino-terminated APTES monolayer and leave at the same time uncovered regions of bare silicon oxide on the substrates. The extreme flatness (rms roughness below 0.5 nm) allows for a highly sensitive monitoring of growth processes by AFM. Together with the chemical variability *Polymer Blend Lithography* (PBL) can become an important tool for studying surface-initiated processes. The quasi two-dimensional chemical patterns open potentials for the application as templates for subsequent self-assembly of inorganic materials, laterally controlled dewetting, constructive lithography or for cell adhesion studies.

## Acknowledgments

The authors thank the Baden-Württemberg Stiftung, for financial support within the Network of Excellence "Functional Nanostructures". Part of the work was supported by the Deutsche Forschungsgemeinschaft (DFG) within the Center for Functional Nanostructures (CFN). Manuel R. Gonçalves, Anne-Marie Saier and Monika Asbach provided substrates. Ulla Nolte took care oft he cell culture. Electron microscopy was provided by Paul Walther and his institute as well as by Anke Leitner.

## References

[1]  J. Robertus, W.R. Browne, and B. L. Feringa. Chem. Soc. Rev., The Royal Society of Chemistry **39**, 354–378 (2010)

[2]  L. G. Parkinson, N. L. Giles, K. F. Adcroft, M. W. Fear, F. M. Wood, and G. E. Poinern. Tissue Engineering Part A **15**, 3753–3763 (2009)

[3]  C. Huang, M. Moosmann, J. Jin, T. Heiler, S. Walheim, and T. Schimmel. *Polymer blend lithography: A versatile method to fabricate nanopatterned self-assembled monolayers*, Beilstein Journal of Nanotechnology **3**, 620–628 (2012)

[4]  L. P. Bauermann, P. Gerstel, J. Bill, S. Walheim, C. Huang, J. Pfeifer, and T. Schimmel. *Templated Self-Assembly of ZnO Films on Monolayer Patterns with Nanoscale Resolution*, Langmuir **26** (6), 3774–3778 (2010)

[5]  J. Sagiv. Journal of the American Chemical Society **102** (1), 92–98 (1980)

[6]  E. Sabatani, I. Rubinstein, R. Maoz, and J. Sagiv. Journal of Electroanalytical Chemistry **219** (1–2), 365–371 (1987)

[7]  A. Ulman. Chemical Reviews **96** (4), 1533–1554 (1996)

[8]   A. Zeira, J. Berson, I. Feldman, R. Maoz, and J. Sagiv. Langmuir **27** (13), 8562–8575 (2011)

[9]   N. K. Chaki and K. Vijayamohanan. Biosensors & Bioelectronics **17** (1–2), 1–12 (2002)

[10]  A. Weddemann, I. Ennen, A. Regtmeier, C. Albon, A. Wolff, K. Eckstaedt, N. Mill, M. K. H. Peter, J. Mattay, C. Plattner, N. Sewald, and A. Huetten. Beilstein Journal of Nanotechnology **1**, 75–93 (2010)

[11]  N. Faucheux, R. Schweiss, K. Lutzow, C. Werner, and T. Groth. Biomaterials **25** (14), 2721–2730 (2004)

[12]  A. Gölzahäuser, W. Eck, W. Geyer, V. Stadler, T. Weimann, P. Hinze, and M. Grunze. Advanced Materials **13** (11), 806–809 (2001)

[13]  Z. She, A. DiFalco, G. Haehner, and M. Buck. Beilstein Journal of Nanotechnology **3**, 101–113 (2012)

[14]  R. Garcia and R. Perez. Surface Science Reports **47** (6–8), 197–301 (2002)

[15]  F. J. Giessibl. Reviews of Modern Physics **75** (3), 949–983 (2003)

[16]  W. M. van Spengen, V. Turq, and J. W. M. Frenken. Beilstein Journal of Nanotechnology **1**, 163–171 (2010)

[17]  G. Malegori and G. Ferrini. Beilstein Journal of Nanotechnology **1** 172–181 (2010)

[18]  T. Koenig, G. H. Simon, L. Heinke, L. Lichtenstein, and M. Heyde. Beilstein Journal of Nanotechnology **2** 1–14 (2011)

[19]  S. Magonov and J. Alexander. Beilstein Journal of Nanotechnology **2**, 15–27 (2011)

[20]  T. Glatzel, L. Zimmerli, S. Kawai, E. Meyer, L.-A. Fendt, and F. Diederich. Beilstein Journal of Nanotechnology **2** 34–39 (2011)

[21]  B. Stegemann, M. Klemm, S. Horn, and M. Woydt. Beilstein Journal of Nanotechnology **2**, 59–65 (2011)

[22]  G. Elias, T. Glatzel, E. Meyer, A. Schwarzman, A. Boag, and Y. Rosenwaks. Beilstein Journal of Nanotechnology **2**, 252–260 (2011)

[23]  M. Jaafar, O. Iglesias-Freire, L. Serrano-Ramon, M. Ricardo Ibarra, J. Maria de Teresa, and A. Asenjo. Beilstein Journal of Nanotechnology **2**, 552–560 (2011)

[24]  J. Stadler, T. Schmid, L. Opilik, P. Kuhn, P. S. Dittrich, and R. Zenobi. Beilstein Journal of Nanotechnology **2**, 509–515 (2011)

[25]  X. Zhang, A. Beyer, and A. Goelzhaeuser. Beilstein Journal of Nanotechnology **2**, 826–833 (2011)

[26]  E. Tirosh, E. Benassi, S. Pipolo, M. Mayor, M. Valasek, V. Frydman, S. Corni, and S. R. Cohen. Beilstein Journal of Nanotechnology **2**, 834–844 (2011)

[27]  S. Xu and G. Y. Liu. Langmuir **13** (2), 127–129 (1997)

[28]  R. D. Piner, J. Zhu, F. Xu, S. H. Hong, and C. A. Mirkin. Science **283** (5402), 661–663 (1999)

[29]  K. Salaita, Y. Wang, and C. A. Mirkin. Nature Nanotechnology **2** (3), 145–155 (2007)

[30]  E. Gnecco. Beilstein Journal of Nanotechnology **1**, 158–162 (2010)

[31]  R. Garcia, R. V. Martinez, and J. Martinez. Chemical Society Reviews **35** (1), 29–38 (2006)

[32] S. Darwich, K. Mougin, A. Rao, E. Gnecco, S. Jayaraman, and H. Haidara. Beilstein Journal of Nanotechnology **2**, 85–98 (2011)

[33] C. Obermair, A. Wagner, and T. Schimmel. *The atomic force microscope as a mechano-electrochemical pen*, Beilstein Journal of Nanotechnology **2**, 659–664 (2011)

[34] S. Walheim, M. Boltau, J. Mlynek, G. Krausch, and U. Steiner. *Structure formation via polymer demixing in spin-cast films*, Macromolecules **30** (17), 4995–5003 (1997)

[35] K. Tanaka, A. Takahara, and T. Kajiyama. Macromolecules **29** (9), 3232–3239 (1996)

[36] J. Genzer and E. J. Kramer. Physical Review Letters **78** (26), 4946–4949 (1997)

[37] P. Mansky, Y. Liu, E. Huang, T. P. Russell, and C. J. Hawker. Science **275** (5305), 1458–1460 (1997)

[38] G. Reiter. Physical Review Letters **68** (1), 75–78 (1992)

[39] G. Reiter. Langmuir **9** (5), 1344–1351 (1993)

[40] K. Kargupta and A. Sharma. Physical Review Letters **86** (20), 4536–4539 (2001)

[41] A. Budkowski, A. Bernasik, P. Cyganik, J. Raczkowska, B. Penc, B. Bergues, K. Kowalski, J. Rysz, and J. Janik. Macromolecules **36** (11), 4060–4067 (2003)

[42] S. Walheim, E. Schaffer, J. Mlynek, and U. Steiner. *Nanophase-separated polymer films as high-performance antireflection coatings*, Science **283** (5401), 520–522 (1999)

[43] L.-M. Chen, Z. Xu, Z. Hong, and Y. Yang. Journal of Materials Chemistry **20** (13), 2575–2598 (2010)

[44] B. Schmidt-Hansberg, M. F. G. Klein, K. Peters, F. Buss, J. Pfeifer, S. Walheim, A. Colsmann, U. Lemmer, P. Scharfer, and W. Schabel. *In situ monitoring the drying kinetics of knife coated polymer-fullerene films for organic solar cells*, Journal of Applied Physics **106** (12), (2009)

[45] K.-H. Yim, W. J. Doherty, W. R. Salaneck, C. E. Murphy, R. H. Friend, and J.-S. Kim. Nano Letters **10** (2), 385–392 (2010)

[46] A. Koehnen, N. Riegel, D. C. Mueller, and K. Meerholz. Advanced Materials **23** (37), 4301–4305 (2011)

[47] B. Schmidt-Hansberg, M. Baunach, J. Krenn, S. Walheim, U. Lemmer, P. Scharfer, and W. Schabel. *Spatially resolved drying kinetics of multi-component solution cast films for organic electronics*, Chemical Engineering and Processing **50** (5–6), 509–515 (2011)

[48] M. Pohjakallio, T. Aho, K. Kontturi, and E. Kontturi. Soft Matter **7** (2), 743–748 (2011)

[49] M. Boltau, S. Walheim, J. Mlynek, G. Krausch, and U. Steiner. *Surface-induced structure formation of polymer blends on patterned substrates*, Nature **391** (6670), 877–879 (1998)

[50] P. Andrew and W. T. S. Huck. Soft Matter **3** (2), 230–237 (2007)

[51] L. Cui, Z. X. Zhang, X. Li, and Y. C. Han. Polymer Bulletin **55** (1–2), 131–140 (2005)

[52] M. Ma, Z. He, J. Yang, Q. Wang, F. Chen, K. Wang, Q. Zhang, H. Deng, and Q. Fu. Langmuir **27** (3), 1056–1063 (2011)

[53] S. C. Thickett, A. Harris, and C. Neto. Langmuir **26** (20), 15989–15999 (2010)

[54] S. Y. Heriot and R. A. L. Jones. Nature Materials **4** (10), 782–786 (2005)

[55] J. Zemla, M. Lekka, J. Raczkowska, A. Bernasik, J. Rysz, and A. Budkowski. Biomacromolecules **10** (8), 2101–2109 (2009)

[56] K. Kawamura, K. Yokoi, and M. Fujita. Chemistry Letters **39** (3), 254–256 (2010)

[57] L. Fang, M. Wei, C. Barry, and J. Mead. Macromolecules **43** (23), 9747–9753 (2010)

[58] E. Ferrari, P. Fabbri, and F. Pilati. Langmuir **27** (5), 1874–1881 (2011)

[59] D. U. Ahn and Y. Ding. Soft Matter **7** (8), 3794–3800 (2011)

[60] M. Ma, Z. He, J. Yang, Q. Wang, F. Chen, K. Wang, Q. Zhang, H. Deng, and Q. Fu. Langmuir **27** (3), 1056–1063 (2011)

[61] A. D. F. Dunbar, P. Mokarian-Tabari, A. J. Parnell, S. J. Martin, M. W. A. Skoda, and R. A. L. Jones. European Physical Journal E **31** (4), 369–375 (2010)

[62] L. Fang, M. Wei, C. Barry, and J. Mead. Macromolecules **43** (23), 9747–9753 (2010)

[63] X. Wang, H. Azimi, H.-G. Mack, M. Morana, H.-J. Egelhaaf, A. J. Meixner, and D. Zhang. Small **7** (19), 2793–2800 (2011)

[64] T. Geldhauser, S. Walheim, T. Schimmel, P. Leiderer, and J. Boneberg. *Influence of the Relative Humidity on the Demixing of Polymer Blends on Prepatterned Substrates*, Macromolecules **43** (2), 1124–1128 (2010)

[65] H. Gliemann, A. T. Almeida, D. F. S. Petri, and T. Schimmel. *Nanostructure formation in polymer thin films influenced by humidity*, Surface and Interface Analysis **39** (1), 1–8 (2007)

[66] W. Madej, A. Budkowski, J. Raczkowska, and J. Rysz. Langmuir **24** (7), 3517–3524 (2008)

[67] C.-C. Chang, T.-Y. Juang, W.-H. Ting, M.-S. Lin, C.-M. Yeh, S. A. Dai, S.-Y. Suen, Y.-L. Liu, and R.-J. Jeng. Materials Chemistry and Physics **128** (1–2), 157–165 (2011)

[68] C. K. Saner, K. L. Lusker, Z. M. LeJeune, W. K. Serem, and J. C. Garno. Beilstein Journal of Nanotechnology **3**, 114–122 (2012)

[69] S. H. Choi and B. M. Z. Newby. Langmuir **19** (18), 7427–7435 (2003)

[70] Q. Wan, Q. H. Li, Y. J. Chen, T. H. Wang, X. L. He, J. P. Li, and C. L. Lin. Applied Physics Letters **84** (18), 3654–3656 (2004)

[71] K.-W. Chae, Q. Zhang, J. S. Kim, Y.-H. Jeong, and G. Cao. Beilstein Journal of Nanotechnology **1**, 128–134 (2010)

[72] E. R. Waclawik, J. Chang, A. Ponzoni, I. Concina, D. Zappa, E. Comini, N. Motta, G. Faglia, and G. Sberveglieri. Beilstein Journal of Nanotechnology **3**, 368–377 (2012)

[73] R. Maoz and J. Sagiv. Advanced Materials **10** (8), 580 (1998)

[74] Y. Paz. Beilstein Journal of Nanotechnology **2**, 845–861 (2011)

[75] C. Wu, Q. Cheng, S. Sun, and B. Han. Carbon **50** (3), 1083–1089 (2012)

[76] R. Sherman, D. Hirt, and R. Vane. Journal of Vacuum Science & Technology A–Vacuum Surfaces and Films **12** (4), 1876–1881 (1994)

[77]  G. Binnig, C. F. Quate, and C. Gerber. Phys. Rev. Lett. **56**, 930 (1986)

[78]  G. S. Shekhawat and V. P. Dravid. Science **310**, 89 (2005)

[79]  H. J. Butt, E. K. Wolff, S. A. C. Gould, B. D. Northern, C. M. Peterson, and P. K. Hansma. J. Struct. Biol. **105**, 54 (1990)

[80]  M. Fritz, M. Radmacher, and H. E. Gaub. Biophys. J. **66**, 1328 (1994)

[81]  M. Radmacher, M. Fritz, and P. K. Hansma. Biophys. J. **69**, 264 (1995)

[82]  H. J. Butt. Biophys. J. **60**, 777 (1991)

[83]  J. H. Hoh, J. P. Cleveland, C. B. Prater, J. P. Revel, and P. K. Hansma. J. Am. Chem. Soc. **114**, 4917 (1992)

[84]  H. J. Butt, B. Cappella, and M. Kappl. Surf. Sci. Rep. **59**, 1 (2005)

[85]  D. Wirtz. Annual Review of Biophysics **38**, 301–326 (2009)

[86]  A. Leitner, T. Paust, O. Marti, P. Walther, H. Herrmann, and M. Beil: *Properties of intermediate filament networks assembled from keratin 8 and 18 in the presence of Mg$^{2+}$*. Biophysical Journal **103**, 195–201 (2012)

[87]  M. Sailer, K. Höhn, S. Lück, V. Schmidt, M. Beil, and P. Walther: *Novel electron 12 tomographic methods to study the morphology of keratin filament networks*. Microsc. Microanal. **16**, 462–71 (2012)

[88]  R. Fürth. Z. Physik **2**, 244 (1920)

[89]  C. Tsallis. J. Stat. Phys **52**, 479 (1988)

[90]  C. Tsallis, J. Stat. Phys. **102**, 259–283 (2001)

[91]  R. Silva, A. R. Plastino, and J. A. S. Lima. Phys. Lett. A **249**, 401 (1998)

[92]  A. Upadhyaya, J. P. Rieu, J. A. Glazier, and Y. Sawada. Physica A **293**, 549 (2001)

[93]  C. Tsallis, S. V. F. Levy, A. M. C. Souza, and R. Maynard. Phys. Rev. Lett. **75**, 3589 (1995)

[94]  A. Compte. Phys. Rev. E **53**, 4191 (1996)

[95]  C. Tsallis and D. J. Bukman. Phys. Rev. E **54**, R2197 (1996)

[96]  L. Borland. Phys. Rev. E **57**, 6634 (1998)

[97]  A. Compte and D. Jou. J. Phys. A **29**, 4321 (1996)

[98]  D. H. Zanette and P. A. Alemany. Phys. Rev. Lett. **75**, 366 (1995)

[99]  C. M. Lo, H. B. Wang, M. Dembo, and Y. Wang. Biophys. J. **79**, 144 (2000)

[100]  N. Zaari, P. Rajagopalan, S. K. Kim, A. J. Engler, and J. Y. Wong. Adv. Mater. **16**, 2133 (2004)

[101]  A. J. Turski, B. Atamaniuk, and E. Turska. J. Tech. Phys. **44**, 193 (2003)

[102]  I. M. Sokolov and J. Klafter. Phys. Rev. Lett. **97**, 140602 (2006)

[103]  H. Keller and P. G. Debrunner. Phys. Rev. Lett., **45**, 68 (1980)

[104]  R. Wu and K. M. Berland. Biophys. J. **95**, 2049 (2008)

[105]  W. Chen, H. G. Sun, X. D. Zhang, and D. Korosak. Comput. Math. Appl. **59**, 1754 (2010)

[106]  G. E. Uhlenbeck and L. S. Ornstein. Phys. Rev. **36**, 823 (1930)